\documentclass[12pt,thmsa]{article}
\usepackage{amssymb}

\usepackage{sw20aip}

\input{tcilatex}
\begin{document}

\author{\qquad \qquad Ioan Sturzu, Physics Department, \and ''Transilvania''
University, Brasov, Romania}
\title{Unsharp measurements and the conceptual problems of Quantum Theory}
\date{May, 1999 }
\maketitle

\begin{abstract}
The paper emphasis the role of unsharpness in the body of Quantum Theory and
the relations to the conceptual problems of the Theory

Key words: quantum measurement, unsharpness, effect, positive
operator-valued measure
\end{abstract}

In a letter to Heisenberg (1921) Pauli wrote:

''One may view the world with a p-eye (an eye for the momentum space) and
one may view it with a q-eye (an eye for the position space), but if one
opens both eyes simultaneously, one than gets crazy.''

This challenging affirmation is characteristic for a whole ideology raised
on the structure of quantum mechanics. However, nowadays, this became an
old-fashioned ideology; thus, one may reformulate Pauli sentence as:

''One may \textbf{imagine} viewing the world with a p-eye and one may 
\textbf{imagine} viewing it with a q-eye, but if one \textbf{imagine}
opening both eyes simultaneously, one than \textbf{imagine} craziness.''

\section{The outline of a formal theory of measurement}

There are some primitive concepts one need when setting up a formal theory
of experimental-type measurement [1], i.e., system, state, property, etc.
Given a property $p$ one can define a counterproperty $\tilde{p}$, which is
false when $p$ is true and is true when is false. The counterproperty has to
be unique. Always one can define two trivial properties: the truistic one $I$
and the impossible one $O$. Of course, $O=\tilde{I}$. A non-trivial property
is sharp if $p\wedge \tilde{p}=O$ ($\wedge $ is logical conjunction).$I$ and 
$O$ are sharp properties. If a property is not sharp, it is unsharp. If a
property is either true or false in a given state, it is real in that state.

If a collection of properties completely defines a state, it is a complete
observable. Two properties are coexistent if there is a complete observable,
which contains both of them. A collection of coexistent properties defines a
(simple) observable. A measurement is a process used for identifying the
state of a system using a complete observable. If all the properties that
belong to a complete observable are sharp, its measurement is a sharp one.
Two observable are coexistent if there exist a complete observable, which
contains both of them.

\section{Classical mechanics}

In classical mechanics, the properties are Borel subsets of the phase space $%
\Omega $, while states are (normalized) measurable functions on $\Omega $
and an observable is a real-function $A:\Omega \rightarrow \mathbf{R}$, or,
equivalently, the set-function induced by it on the set of properties. Using
infinite series of properties and a frequencial definition of the state [1]
one may identify the upper definition of a complete observable. In classical
mechanics all properties are sharp (the counterproperty is the
set-complement of the corresponding Borel set), and all properties are
mutually coexistent (the presence of a Borel set in the infinite series is
unimportant).

The concept of ''system'' does not refer mere to a single physical object
(or a collection of objects), but to a countable ensemble of identical
copies of one single object, while measurement is determination of the
normalized measurable function frequencially defined in that ensemble.

Of course, these are mere ideal definitions; practically one uses finite
number of copies. Nevertheless, the properties defining the state identified
by the measurement can be fully assign to any of the individual physical
object from the ensemble. Note that one may identify singularly
distributions as solutions for the state functions (i.e. the system has a
definite position, etc.), but these represent mere ideal cases whose meaning
is that one may increase unlimitedly the precision of measurement.

\section{Quantum mechanics in the sharp measurement interpretation}

Quantum mechanics needed another mathematical representation from the point
of view of measurement formalism. The states are represented by positive
trace-class operators $\rho $ on a separable Hilbert space $\mathcal{H}$,
while properties seemed to be represented by projector operators $p$ on the
closed subspaces of $\mathcal{H}$. Defining the counterproperty $\tilde{p}$
as the orthogonal complement $p^{\bot }$ and identifying logical conjunction
of properties with the projector on the intersection of the corresponding
subspaces, one may see that all these properties are sharp ones: $p\wedge 
\tilde{p}=O$ [2].

For any state there are many self-adjoint operators, that may define it
completely. By spectral theorem, to these self-adjoint operators correspond
spectral measures whose ranges are the collections of properties
(projectors) which, by the upper definition, defines an observable. Two
projectors (sharp properties) are coexistent if and only if they commute: 
\[
p_1\cdot p_2=p_2\cdot p_1 
\]

Also, in quantum mechanics one has to define the system as an ensemble of
identically prepared physical (micro-)objects. For any complete observable
there exist an experimental procedure which assigns to every projector from
the range of the spectral measure a filter, so that an individual object,
interacting with it, will pass or not; consequently the corresponding sharp
property would be true, or false. Note that for another individual object
the answer need not be the same; when the answer is always true or always
false the property is said to be real for that state.

Some projector $P_1$ may not belong to the range of the spectral measure of
any other complete observable where $P_2$ is present. So, there exist
non-coexistent properties in quantum mechanics. The filters corresponding to
such non-coexistent properties cannot be built up together. This is the case
of properties related to non-commuting operators, known as complementary
observables. Consequently, the experimental procedures corresponding to two
sharp complementary observables cannot be built up together simultaneously.
Heisenberg had been the first who investigated the result of two consecutive
non-coexistent measurements, and obtained an euristical relation (related to
a similar one from the wave theory), which sounds like a relation obtained
by independent means from the mathematical formalism of the quantum theory.
The first is known as Heisenberg relation, while the later as Robertson's
[3]: 
\[
\Delta A\cdot \Delta B\geqslant \frac \hbar 2 
\]

This similarity was, at that time, the single reason to assign the
significance of imprecision in measuring an observable to the second order
symmetric moment of it, i.e. $\Delta A,$ $\Delta B$. Nevertheless, this is
the real significance, but the sharp version of the quantum formalism was
not able to produce the proof, and this was the standpoint for a lot of
conceptual confusions.

Briefly, if one can ''see'' the world only by sharp properties, reality has
to be split in complementary sub-realities, each giving us full access
unless we totally disregard the other sub-realiti(-es). Nevertheless, this
kind of knowledge has to be complete, because one can fully find the state
of the system. This claiming has challenged A. Einstein scientific realism
who proved [4] that (sharp) quantum theory cannot be simultaneously: real,
causal, local and complete[5]. Many scientists interpreted Einstein's
criticism as a conscription for finding a classical-type (hidden-variable)
theory, which was intended to fill the presumable incompleteness of quantum
theory. Nevertheless a theorem due to J.S. Bell [6] followed by experimental
verifications [7] confirmed that predictions of quantum theory for sharp
observables are in perfect concordance with physical reality, contrary to
hidden-variable-classical-type theories.

\section{Quantum mechanics in the unsharp measurement interpretation}

There were many attempts to solve these problems by giving up the reality,
causality or locality conditions, contrary to Einstein views, and to common
scientific sense, too. By the upper presentation, it is clear that one has
another solution for Einstein incompleteness problem, i.e., giving up the
sharpness condition. Indeed, this renunciation is a quite natural one: by
the M. Born interpretation, the sole quantity subject to direct experimental
verification is the probability: 
\[
w=Tr(\rho \cdot p) 
\]

Of course is subject to the condition:

\begin{equation}
0\leq w\leq 1  \label{prob}
\end{equation}

which corresponds in the set of the projectors to:

\begin{equation}
O\leq p\leq I  \label{proiect}
\end{equation}

Relation (2) is a necessary and sufficient condition for (1), but is not a
necessary condition for $p$ to be a projector. Indeed, there exist more
general operators, called effects, which obey relation (\ref{proiect}) and
can stand for generalized properties (E. Davies and C. Helstrom were the
first who introduced [8] this notion. Afterwards, the topic was rigurosly
grounded by G. Ludwig [9], and developed by E. Prugovecki, S. Gudder, etc.).
One may define the counterproperty:

\begin{equation}
\tilde{p}=I-p  \label{comp}
\end{equation}

This is not, generally, an orthocomplement, because the conjunction need not
necessarily exist, and if it exist, it is not necessarily $O$[2]. The
operator $\frac 12\cdot I$ is called semitransparent effect. The effects
which are neither less than the semitransparent effect nor greater than it,
are called regular effects. These regular effects correspond to the most
general notion of property. The complementation (\ref{comp}) is a kind of
orthocomplementation in the set of regular effects, i.e., regular effects
are sharp ones in this sense (of course they are not real sharp ones). Now,
one may define a complete unsharp observable by a collection of effects,
which completely defines the state. Two unsharp properties are coexistent if
there exist a complete observable, which contains both of them (of course,
it has to be an unsharp one). In general, there does not exist any
self-adjoint operator corresponding to an unsharp observable. Indeed,
instead of the spectral (projectorial) measure of the sharp observable, here
one has a positive- oprerator-valued measure (POVM), which cannot define, by
itself, a self-adjoint operator, but only a maximal symmetric operator.

A POVM F can be obtained by smearing a projectorial measure E: 
\[
F(X)=\int_\Omega w(X,\lambda )\cdot dE_\lambda 
\]
where $\Omega $ is the space of the measurement outcomes, X is a Borel set
on and is a measurable function on for fixed X and a probability measure on
the Borelians class for fixed $\lambda $ [2] . The smearing operation
induces some degree of uncertainty, which is due to the absence of sharply
defined criteria of ascribing numerical values to the equivalence classes
produced by measurement process. Also, POVM may come out in the theory of
open systems and, consequently, in the theory of quantum measurements.

A theorem due to Neimark ensures the extension of any closed maximal
symmetric $A$ operator acting on $\mathcal{H}$ to a selfadjoint one acting
on $\mathcal{\tilde{H}\supset H}$ and, consequently, of any POVM to a
projectorial one. However, this extension is not unique, which means that
one cannot state the sharp version of Quantum Theory as a fundamental one,
able to produce, via smearing processes, particular unsharp cases. By the
contrary, reality is itself fundamentally unsharp, while the sharp
situations are available only for gedanken-experiments.

The conceptual problems of Quantum Theory are, most of all, related to such
gedanken-experiments. The genuine role of such ''experiments'' was to
clarify the concepts of the new theory vs. those of the classical mechanics.
(The later have been constructed by similar procedures; e.g., the idea of a
material point moving without constraints comes from a gedanken-experiment
invented by Galilei). Unfortunately, as we shown upwards, the expected
conceptual clarification did not come.

It was stated upwards that from the formal theory of (experimental)
measurements one yields the statistical character of the concept of system.
In Classical Physics this does not make any harm, because all definable
properties are mutually coexistent, so one can fully assign them to any
individual object described by the system. By the contrary, in Quantum
Physics this cannot be done, so its concepts seem to bear a kind of
incompleteness. Unfortunately, there exist a great reluctance against the
idea of incompleteness of the scientific knowledge: a good theory has to
offer a complete knowledge, even at risk of changing the meaning of the word
''complete''. This reluctance is closely related to the deep origins of the
modern (experimental) physics, which stays in the Middle Age occidental
monasteries. Their new spiritual perspective, contrary to the Orthodox
Christian one, proclaimed (implicitly), for the first time in the human
history, a kind of ''God's death'': God is far away, and the world is all
ours! So, we can handle it as we want, because God let us; from here, one
yields the inquisitorial method, which later became the experimental method:
the subject (of knowledge) owns a kind of pre-knowledge (some information
about some witchcraft vs. a theory to be tested), while the object (of
knowledge) has to obey to rules of the subject, i.e. to answer his
pre-conceived questions (what kind of witchery he had done vs. what definite
property it posses). Unsharpness is a way of sparing the completeness taboo,
i.e. a way which preserves the experimental method (restricting to regular
effects is very important!). Nevertheless, one may conceive a Physics beyond
the experimental method, which is David Bohm's approach to a hidden variable
theory [10]. But, unfortunately, Bohm's approach is undermined by the same
sin as the experimentist approach [11]. The latter imposes a verifiable
unsharp structure to reality, but the method itself is not
(meta-)verifiable, while the former tries to change the method (by changing
the immutable linguistic form of Quantum Theory as a universal theory),
making it (meta-)verifiable, but plays, unavoidable, with imposed
unverifiable sharp structures. 

The Occam's razor tells us that unsharp measurement interpretation of
quantum theory is, for the moment, the most appropriate way to understand
fundamental physics these days.

\bigskip 

Bibliography:

[1] Sturzu, I. - Outlines for a formal theory of physical measurements, to
appear

[2] Busch, P., Grabowski M., Lahti P. - Operational Quantum Physics,
Springer-Verlag, Berlin, 1997

[3]Robertson, H. - Phys.Rev.34,163,1929

[4] Einstein, A., Podolsky B., Rosen N. - Phys.Rev. 47, 778, 1935

[5]P\^{a}rvu, I. - Arhitectura existentei, Bucuresti,1990

[6] Bell, J. S. - Physica 1, 195, (1964)

- Rev.Mod.Phys. 38, 447, (1966)

[7] Aspect, Grangier, Dalibard, Roger - Phys.Rev.Lett. 47, 460, (1981)/49,
91, (1982)

[8] Davies, E. - Quantum Theory of Open Systems, Academic, N.Y. 1976

Helstrom, C. - Quantum Detection and Estimation Theory, Academic, N.Y. 1976

[9]Ludwig, G. - Foundations of Quantum Mechanics, Springer-Verlag, Berlin,
1983

[10]Bohm, D., Bub, J. - Rev.Mod.Phys. 18,3, (1966)

[11]Sturzu, I. - Why are hidden variables so hidden?, to appear

\end{document}